\documentclass{ws-procs11x85}
\usepackage{ws-procs-thm}           
\usepackage{booktabs}
\usepackage{xcolor}
\usepackage{enumitem}
\usepackage{hyperref}
\usepackage{pdfpages}

\setlist[itemize]{noitemsep, topsep=0pt, parsep=0pt, partopsep=0pt}

\begin{document}
\title{Discovery of Disease Relationships via Transcriptomic Signature Analysis Powered by Agentic AI}

\author{Ke Chen}

\address{School of Information Sciences,\\ University of Illinois Urbana-Champaign,\\
Urbana, IL, USA\\
E-mail: kec10@illinois.edu}

\author{Haohan Wang}

\address{School of Information Sciences,\\ University of Illinois Urbana-Champaign,\\
Urbana, IL, USA\\
E-mail: haohanw@illinois.edu}

\begin{abstract}
\textbf{Abstract.} Modern disease classification often overlooks molecular commonalities hidden beneath divergent clinical presentations. This study introduces a transcriptomics-driven framework for discovering disease relationships by analyzing over 1,300 disease–condition pairs using GenoMAS, a fully automated agentic AI system. Beyond identifying robust gene-level overlaps, we develop a novel pathway-based similarity framework that integrates multi-database enrichment analysis to quantify functional convergence across diseases. The resulting disease similarity network reveals both known comorbidities and previously undocumented cross-category links. By examining shared biological pathways, we explore potential molecular mechanisms underlying these connections—offering functional hypotheses that go beyond symptom-based taxonomies. We further show how background conditions such as obesity and hypertension modulate transcriptomic similarity, and identify therapeutic repurposing opportunities for rare diseases like autism spectrum disorder based on their molecular proximity to better-characterized conditions. In addition, this work demonstrates how biologically grounded agentic AI can scale transcriptomic analysis while enabling mechanistic interpretation across complex disease landscapes. All results are publicly accessible at \url{github.com/KeeeeChen/Pathway_Similarity_Network}.
\end{abstract}

\keywords{Disease similarity network; Transcriptomic associations; AI-driven discovery}


\section{Introduction}\label{intro}
Modern disease classification is predominantly grounded in clinical symptoms, anatomical locations, and observable phenotypes\cite{cms2023icd10cm, fauci2008harrison, goh2007disease}. While practical for diagnosis and treatment, this symptom-centric taxonomy often obscures deeper biological relationships between diseases—especially those with divergent clinical manifestations but shared molecular origins\cite{prieto2021nosology, zanin2018heterogeneity}. In contrast, transcriptomic signatures\cite{ferraro2020transcriptomic} capture gene expression patterns directly reflective of underlying cellular mechanisms, offering a biologically principled lens to reexamine disease relationships.

Recent studies have shown that transcriptomic profiling not only reveals disease-specific pathways related to susceptibility \cite{wang2024sle, shao2024itp}, progression\cite{poon2023mmscs, wang2022alzheimers, mendelsohn2024tb}, and resilience\cite{chaudhuri2025vascular, pelaia2023critical}, but also uncovers shared molecular programs\cite{ferraro2020transcriptomic, antcliffe2019vanish, bigot2016cd} across phenotypically distinct diseases. These shared patterns, often invisible to clinical observation\cite{figueiredo2021coexpression, jha2022cancer}, have profound implications for disease reclassification\cite{harrill2024toxscreening, kim2024rcc}, biomarker discover\cite{namba2022target, zhai2024dgsignatlas}, and therapeutic repurposing\cite{lessard2024multiomics, wang2024copd}. However, realizing these benefits at scale remains challenging\cite{anaconda2020datascience, bpc2023intersect}: each transcriptomic dataset requires extensive preprocessing, normalization, and analysis—an effort that is labor-intensive and difficult to replicate consistently across diverse biological and demographic contexts.

To address this, we leveraged GenoMAS\cite{liu2025genomas}, a fully automated, agentic AI system that performs large-scale transcriptomic analyses across 1,384 disease–condition pairs drawn from the GenoTEX\cite{liu2024genotex} benchmark dataset. Each \emph{pair} represents a disease under a specific biological or demographic condition (e.g., age, sex, obesity, comorbidity), enabling nuanced profiling across 132 diseases and 911 cohorts. Powered by a team of specialized LLM agents, the agentic system performs end-to-end processing, from data cleaning to statistical inference, to identify the genes associated with the disease status under the conditions. In other words, the agentic system identified the \emph{transcriptomic signatures} for each \emph{pair} of disease and condition. 

Building on these results, we construct a disease relation network through transcriptomic signatures, identifying statistically significant transcriptomic overlaps between thousands of disease–condition pairs. We validate this network against ICD-10-CM categories and observe both strong within-category clustering and biologically plausible cross-category links—highlighting hidden disease relationships overlooked by traditional taxonomy.

To further interpret the functional basis of these relationships, we extend our analysis to the pathway level. By conducting multi-database enrichment and introducing a novel pathway-based similarity scoring framework, we identify over 1,000 disease combinations that converge on shared molecular pathways. These shared pathways reveal fundamental biological mechanisms that transcend clinical presentation and reflect the cellular logic underlying diverse disease states. 

\begin{figure}
    \centering
    \includegraphics[width=0.95\linewidth]{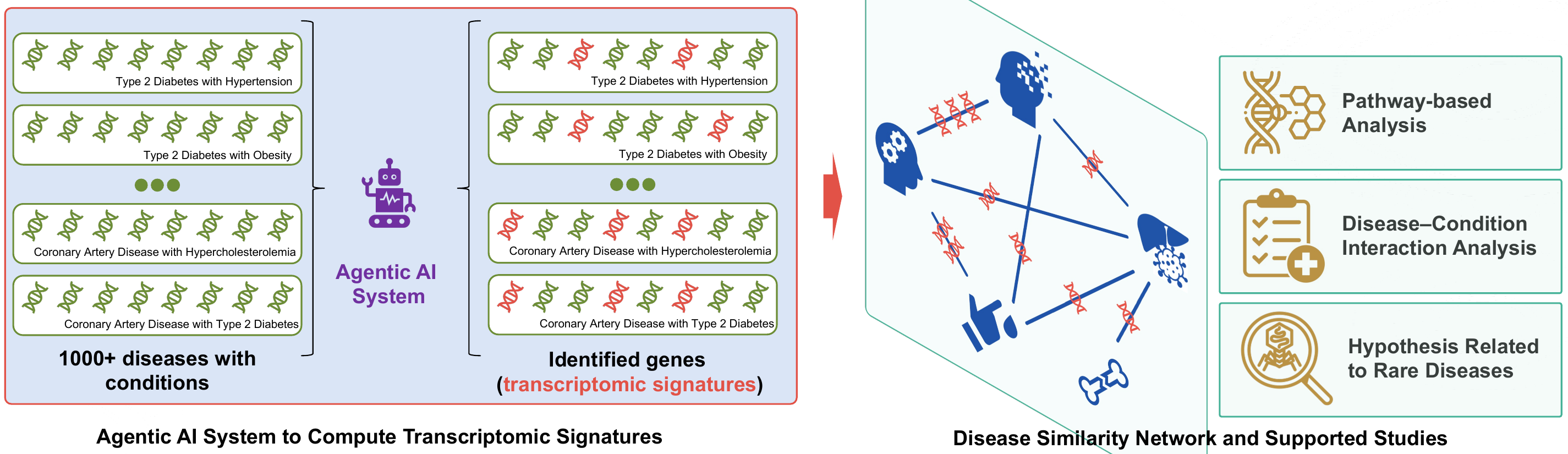}
    \caption{Agenetic AI analysis of transcriptomic data for transcriptomic signatures and the network of diseases constructed from the signatures.}
    \label{fig:main}
\end{figure}

Our analysis recovers well-established comorbidities (e.g., epilepsy and Canavan disease), confirms mechanistically plausible cross-category relationships (e.g., ankylosing spondylitis and osteoporosis), and—most notably—uncovers novel disease links that have not been previously reported in the literature. For these unexpected pairs, we hypothesize potential biological mechanisms supported by shared pathways and gene functions, providing initial interpretive insights to be explored in future studies—for instance, immune and glycosylation-related convergence between Gaucher disease and kidney cancer, or shared metabolic signaling and oxidative stress patterns observed in neurodegenerative diseases and \texttt{Ocular Melanoma}.

Finally, we explore how background conditions like obesity and hypertension modulate disease–disease transcriptomic similarity and highlight rare disease cases, such as autism spectrum disorder, where shared molecular signatures with more common diseases may inform drug repurposing opportunities.

To encourage broader exploration of hidden disease relationships, we have made our full results publicly available at \url{github.com/KeeeeChen/Pathway_Similarity_Network}. Additionally, an initial biological plausibility assessment was conducted using GPT-4o to highlight approximately 200 disease-combinations that exhibit interpretable functional convergence. We hope this resource can inspire new hypotheses and offer alternative perspectives for understanding disease mechanisms beyond established taxonomies.

In summary, the contributions of this paper are illustrated in Figure~\ref{fig:main} and as follows: 
\begin{itemize}
    \item We perform large-scale transcriptomic signature analysis across 1,384 disease-condition pairs using an agentic AI system (GenoMAS).
    \item We construct a gene-level transcriptomic similarity network based on transcriptomic signatures, revealing both strong within-category and cross-category connections.
    \item We introduce a pathway-level similarity framework based on multi-database enrichment and joint pathway scoring, identifying over 1,000 disease-condition combinations that converge on interpretable molecular mechanisms.
    \item We highlight examples of transcriptomic convergence in both well-established and unexpected disease pairings, including several cases with no previously documented clinical or molecular connection.
    \item We study how background conditions such as obesity and hypertension modulate transcriptomic similarity between diseases, and identify rare diseases whose molecular profiles suggest potential therapeutic strategies based on cross-disease alignment.
\end{itemize}


\section{Results}

Before presenting our main findings, we first clarify several key terms used throughout the analysis. Our study involves multiple levels of comparison across diseases, biological conditions, and their combinations. Table~\ref{tab:terminology} summarizes the terminology used.

\begin{table}[htbp]
\footnotesize
\tbl{Terminology used in this study}{
\begin{tabular}{@{}lp{13.5cm}@{}}
\toprule
\textbf{Term} & \textbf{Definition / Example} \\
\midrule
\textbf{Disease} & A clinical diagnosis or condition label. 
\textit{e.g., Liver Cancer} \\
\textbf{Condition} & A biological or demographic modifier that contextualizes the disease. \\
& \textit{e.g., Obesity, Sex, Age, Hypertension} \\
\textbf{Pair} & A disease combined with a specific condition. 
\textit{e.g., Liver Cancer--Obesity} \\
\textbf{Combination} & A pairwise comparison between two disease--condition pairs. \\
& \textit{e.g., (Liver Cancer--Obesity) vs. (Schizophrenia--Gender)} \\
\bottomrule
\end{tabular}
}
\label{tab:terminology}
\end{table}

\subsection{Gene-Based Similarity Network}

To investigate inter-disease relationships at the transcriptomic level, we preliminarily analyzed the statistical significance of overlap of transcriptomic signatures between every \emph{combination} of the 1,384 disease–condition pairs (hereafter, \textit{``combinations"}).

\begin{figure}[htbp]
    \centering
    \includegraphics[width=1\textwidth]{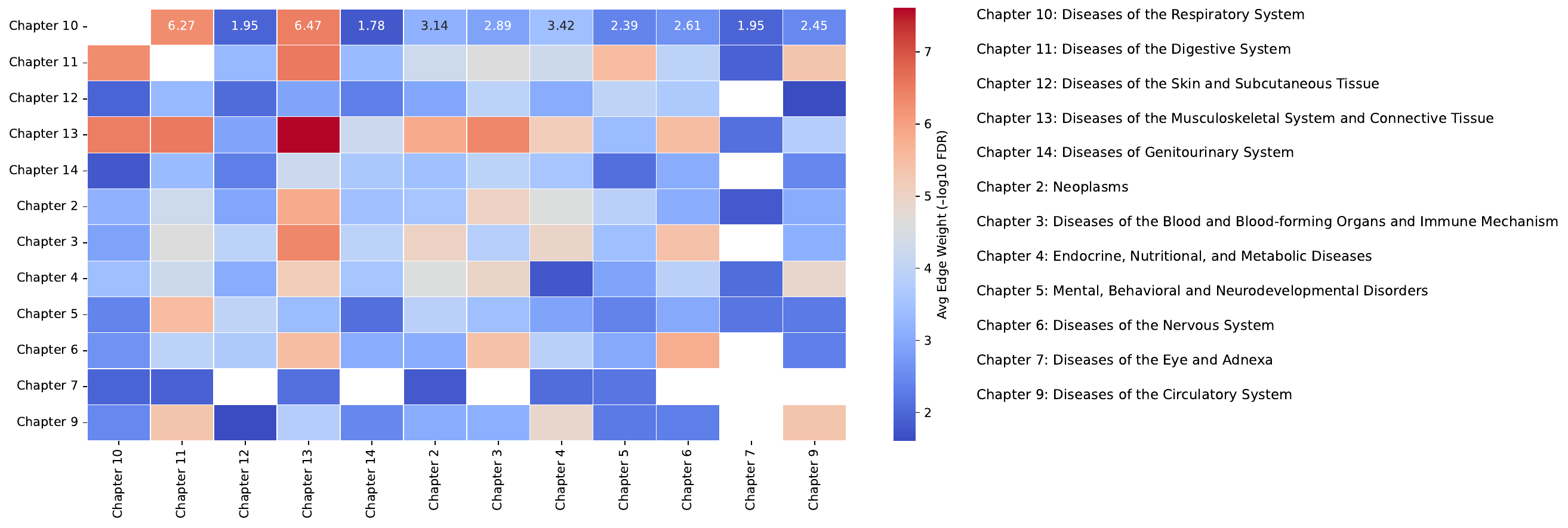}
    \caption{
        Heatmap of average gene-based similarity between ICD-10-CM chapters. 
        Diagonal blocks (e.g., Chapters 6, 9, 13) show strong within-category similarity, 
        while several off-diagonal blocks indicate cross-category transcriptomic convergence. 
        Notably, chapter 2, 3, 10, and 11 show 
        higher similarity to other categories than within their own, suggesting latent biological overlap.
    }
    \label{fig:heatmap}
\end{figure}

Based on these shared gene relationships, we constructed a graph in which each node represents a disease–condition pair, and each edge connects two pairs that significantly share a set of genes (see Section ~\ref{method:gene_similarity} for details). To validate the biological plausibility of the resulting network, we compared our result with the ICD-10-CM classification system \cite{cms2023icd10cm}. Specifically, we prompted GPT-4o to assign an ICD category to each disease, and constructed a heatmap of average pairwise gene similarity scores for both pairs within the same ICD category and cross-category pairs. (Figure~\ref{fig:heatmap}) 

As expected, many chapters show the strongest similarity within their own category—e.g., Chapter 6, 13, and 9 all display elevated diagonal values. However, the heatmap also reveals that several chapters exhibit their highest similarity scores with other categories rather than their own. For instance, certain subtypes within Chapter 2 and 3 share stronger transcriptomic profiles with Chapter 13 than within their own chapters, suggesting biologically meaningful cross-category overlap. While some of these connections may arise from annotation bias or shared tissue origin, others may reflect previously overlooked biological commonalities.

Together, these findings suggest that while disease taxonomy based on anatomy or symptoms often aligns with molecular signatures, gene-level similarity can also uncover latent biological relationships that transcend clinical classifications. This motivated our subsequent pathway-level analysis to probe deeper into shared mechanisms.



%

\subsection{Pathway-Based Disease Similarity}

\begin{figure}[htbp]
\centering
\includegraphics[width= 1\textwidth]{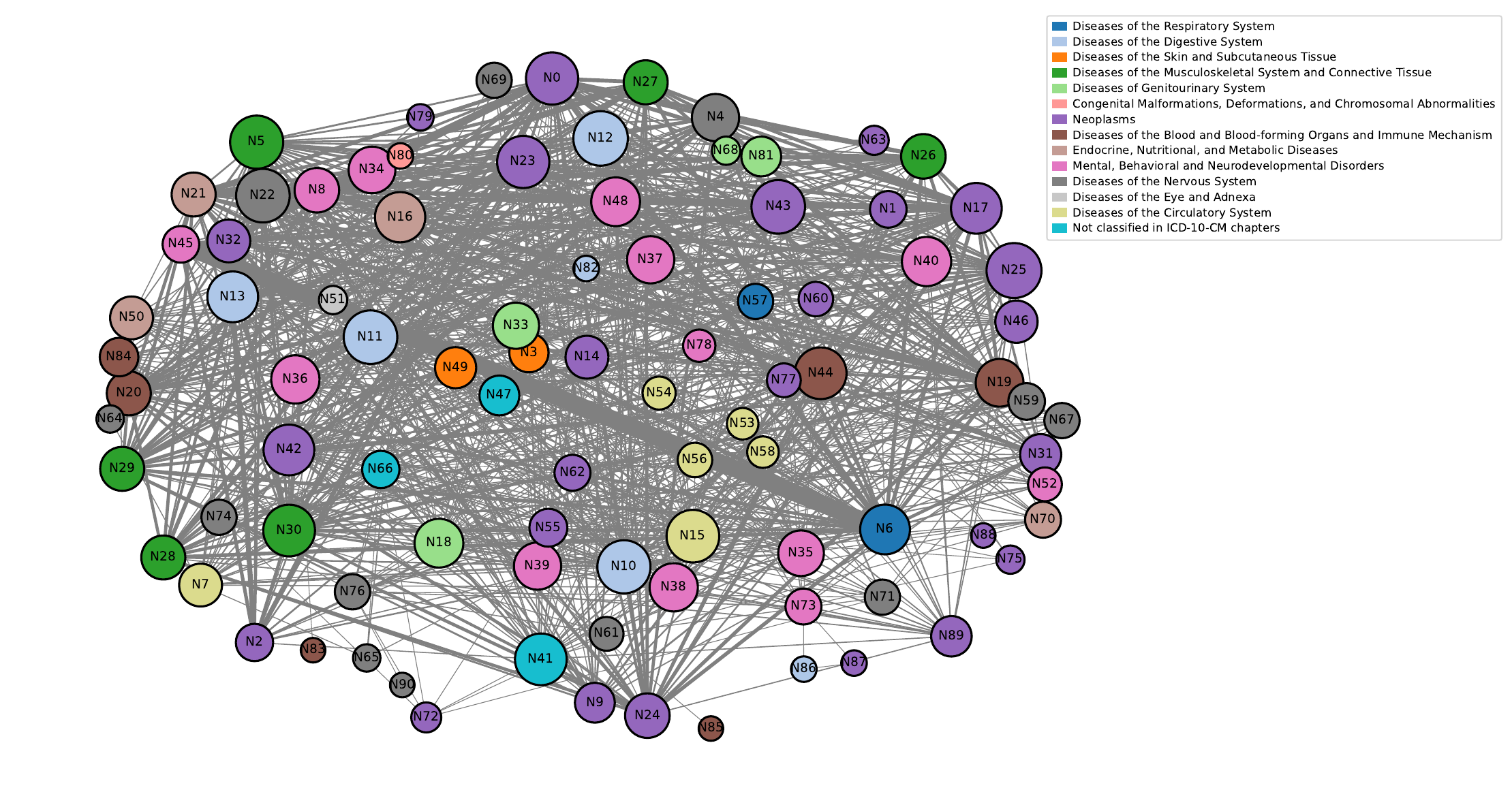}
\caption{Pathway-level similarity network. Each node represents a disease--condition pair, colored by ICD-10-CM category. Edges indicate statistically significant overlap in enriched pathways. Both the thickness and the length of each edge reflect the strength of similarity—stronger pathway-level similarity results in shorter and thicker edges. Node size reflects degree centrality. While many nodes are connected, this visualization is designed to emphasize the strength of similarity rather than the presence of connection.
}

\label{fig:pathway_net}
\end{figure}

While these combinations significantly shared some genes, their biological relevance remained unclear without understanding what molecular processes these genes are involved in. To better interpret the functional basis of disease similarity, we examined pathway-level overlap among the 1,293 significant disease–condition combinations identified by our gene-based analysis (see Section~\ref{method:pathway_similarity} for details). Among these combinations, 1,060 were found to share at least one enriched pathway. To visualize these relationships, we constructed a weighted undirected graph (see Figure~\ref{fig:pathway_net}) to provide a systems-level view of transcriptomic convergence across diseases.

 This network reveals a clear tendency for nodes of the same ICD-10-CM category to cluster together, which suggests that our pathway-based analysis, while agnostic to clinical labels, nonetheless recapitulates key elements of traditional disease taxonomy. At the same time, many edges span across categories, hinting at molecular commonalities that transcend existing clinical boundaries.

Subsequent analyses in this study are grounded in this network representation. Specifically, we focus on interpretable subgraphs extracted from the full network—such as highly connected modules, cross-category clusters, and rare disease neighborhoods—to uncover novel patterns of comorbidity, shared vulnerability, and potential therapeutic convergence. This pathway network thus serves as the functional scaffold for the biological insights that follow.

\subsubsection{Transcriptomic Similarity Reflects Symptom-Based Taxonomy}

Many high-scoring combinations correspond closely to well-established disease relationships. For example, \texttt{Canavan Disease} and \texttt{Epilepsy}—both neurological disorders—significantly shared pathways such as \textbf{detection of chemical stimulus}, \textbf{sensory perception}, and \textbf{G protein-coupled receptor signaling pathway}. These pathways are central to neuronal communication and signal transduction, especially in sensory and stimulus-related neural activity. 
This is consistent with clinical consensus.

\begin{table}[htbp]
\tbl{P-values of pathological-related shared Pathways in Canavan Disease and Epilepsy}{
\begin{tabular}{lcc}
\toprule
\textbf{Pathway} & \textbf{Canavan\_Disease--None} & \textbf{Epilepsy--None} \\
\midrule
detection of chemical stimulus & $2.04 \times 10^{-22}$ & $3.72 \times 10^{-26}$ \\
sensory perception             & $6.01 \times 10^{-16}$ & $1.50 \times 10^{-22}$ \\
G protein-coupled receptor signaling pathway & $1.17 \times 10^{-13}$ & $2.43 \times 10^{-17}$ \\
\bottomrule
\end{tabular}}
\label{tab:pathway_pvals}
\end{table}

There are also other top-scoring combinations aligned with known biological and clinical groupings, including:
- \texttt{Stomach Cancer} and \texttt{Peptic Ulcer Disease}, both involving the gastrointestinal system;
- \texttt{Depression} and \texttt{Schizophrenia}, both major psychiatric disorders;
- \texttt{Bladder Cancer} and \texttt{Endometrioid Cancer}, which share hormonal and tissue-level commonalities.

\subsubsection{Cross-Category Transcriptomic Similarity with Empirical Support} 
Beyond well-established within-category associations, our pathway-based analysis also revealed biologically meaningful links across some phenotypically unrelated disease categories.

One example is \texttt{Ankylosing Spondylitis (AS)} and \texttt{Osteoporosis}, two conditions traditionally categorized under musculoskeletal and metabolic disorders, respectively. They significantly share genes such as \textit{AAMDC}, \textit{ABCB1}, and \textit{ABCA5}, along with enriched pathways related to lipid metabolism, cholesterol regulation, ABC transporters, steroid biosynthesis, and xenobiotic response.

These functions jointly regulate inflammation, immune activity, and bone remodeling—suggesting a shared biological axis linking chronic inflammation, lipid dysregulation, and bone loss. This supports the hypothesis that inflammatory mechanisms in AS may drive osteoporosis risk through disrupted metabolic signaling. Our findings are consistent with recent empirical studies confirming an elevated osteoporosis risk in AS patients \cite{sharif2023osteoporosis,mei2023mendelian}, and with transcriptomic evidence highlighting immune-driven bone density reduction \cite{zhang2022immune}. Our results further clarify potential shared molecular mechanisms underlying this comorbidity.

We also observed high pathway-based similarity between \texttt{Hemochromatosis} and \texttt{Liver Cancer}, supported by significantly shared genes such as \textit{AADAT}, \textit{A1BG}, \textit{A4GNT}, and \textit{AARS2}. These genes participate in pathways related to amino acid metabolism, mitochondrial function, immune regulation, and glycoprotein processing.

These shared pathways converge on several key processes: iron overload in Hemochromatosis promotes oxidative stress and chronic inflammation in the liver—an organ central to both conditions. Pathways such as \textit{Tryptophan metabolism}, \textit{immune response signaling}, and \textit{protein glycosylation} highlight a potential mechanistic chain involving metabolic disruption, immune imbalance, and epithelial cell proliferation—all of which may facilitate hepatocarcinogenesis.

These findings align with prior epidemiological studies confirming elevated liver cancer risk in patients with HFE-related Hemochromatosis \cite{olynyk2021hemochromatosis}, and extend beyond prior expression analyses by identifying a broader set of molecular mediators \cite{jayachandran2020hhc}.


These examples illustrate how pathway-level similarity can provide complementary context to gene-level overlap, offering candidate functional processes that may help interpret co-occurrence patterns between diseases.

\subsubsection{Transcriptomic Similarities That Are Potentially Unexpected from Conventional View}\label{novel_results}

One interesting outcome of our transcriptomic similarity analysis is the resemblance observed between several phenotypically and clinically unrelated conditions.
One example is \texttt{Gaucher Disease} and \texttt{Kidney Chromophobe}. Although \texttt{Gaucher Disease} is a lysosomal storage disorder and \texttt{Kidney Chromophobe} is a renal carcinoma subtype, they share significant expression of genes such as \textit{A1BG}, \textit{A4GNT}, and \textit{A2M}, alongside co-enrichment in pathways involving immune signaling, extracellular matrix (ECM) remodeling, and protein glycosylation.

These overlapping genes suggest a common functional landscape shaped by immune regulation, protein processing, and inflammation. \textit{A1BG} has been linked to tumor-associated immune modulation\cite{tian2008pancreatic, piyaphanee2011a1bg}; \textit{A4GNT} influences glycosylation—a process central to immune escape and cellular signaling\cite{fujii2022mucin1}; and \textit{A2M} is involved in ECM maintenance and inflammatory control\cite{sun2023a2m}. Pathway-level analysis further reveals enrichment in immune response, ECM organization, glycoprotein biosynthesis, and cellular stress adaptation. Together, these findings point to a shared cellular environment marked by chronic inflammation and metabolic stress—hallmarks of both lysosomal disorders and tumorigenesis. While no direct clinical relationship has been reported between \texttt{Gaucher Disease} and \texttt{Kidney Cancer}, our results suggest a potentially overlooked biological intersection that warrants further investigation.

A second example involves an unexpected transcriptomic connection between \texttt{Alzheimer’s Disease \& Parkinson’s Disease}, and \texttt{Ocular Melanoma}. These conditions share significant expression of \textit{AADAT} and \textit{AASDH}, genes involved in lysine\cite{okuno1993aminotransferase, hallen2013lysine} and tryptophan metabolism\cite{okuno1993aminotransferase}, which regulate $NAD^+$ biosynthesis, glutamate balance, oxidative stress response, and immune modulation\cite{essa2013nad, yang2023amino, michaudel2023tryptophan}. Though these processes diverge in pathological outcomes, they are central to both neurodegeneration and cancer.

\begin{figure}[htbp]
    \centering
    \includegraphics[width=0.8\textwidth]{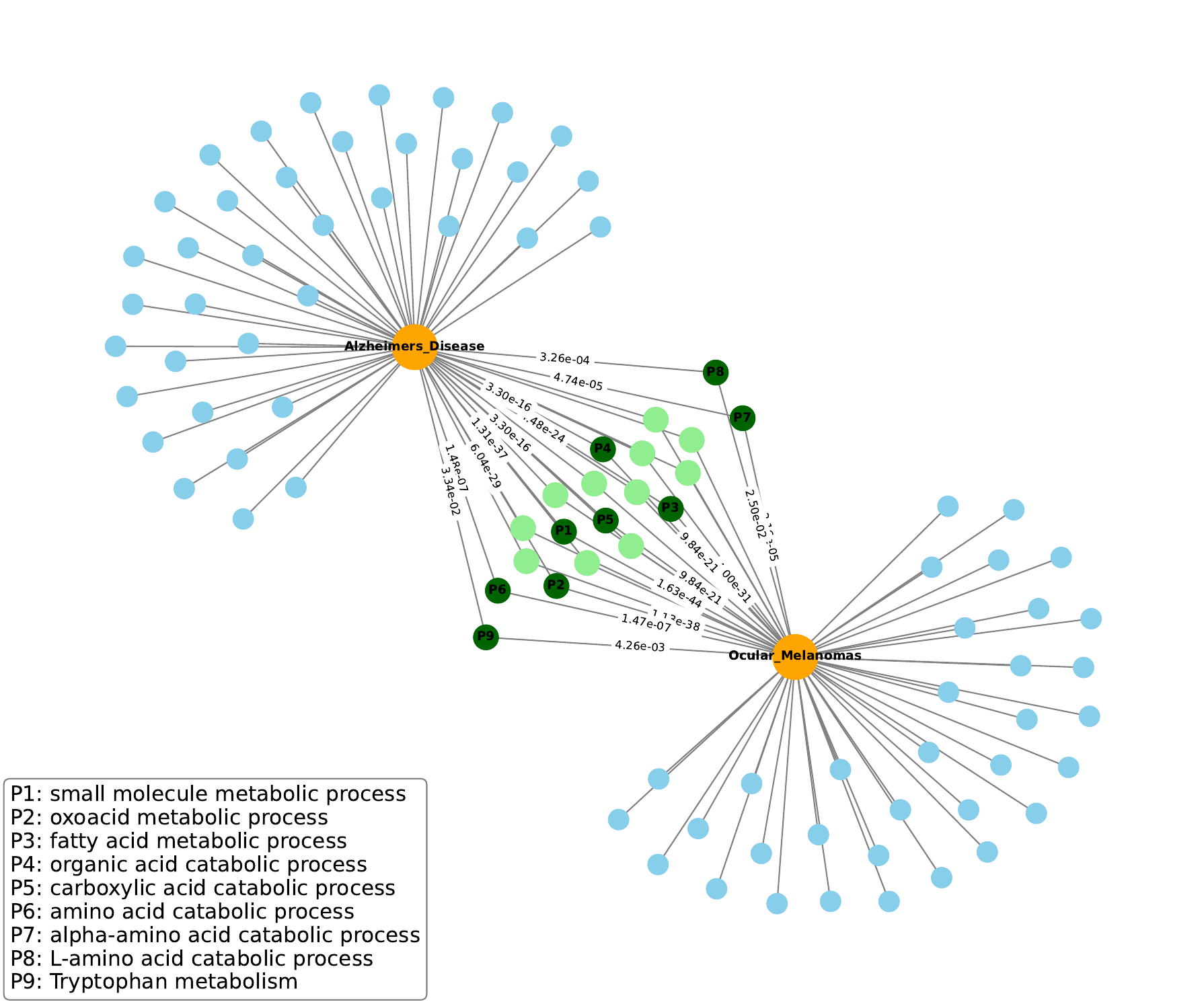}
    \caption{
        Shared transcriptomic pathways between Alzheimer’s Disease and Ocular Melanoma. The graph displays the top 50 most significant enriched pathways in each disease. Blue nodes represent highly enriched but not shared pathways.
        Green nodes indicate pathways shared by both diseases, with darker green highlighting the pathways discussed in~\ref{novel_results}, which are potentially relevant to the comorbidity, such as amino acid catabolism and $\beta$-oxidation (see legend). 
        Edge labels reflect pathway significance (p-values), and edge lengths scale with significance.}
    \label{fig:ad-om-pathways}
\end{figure}

As shown in Figure~\ref{fig:ad-om-pathways}, we observed shared enrichment in pathways related to amino acid catabolism, $\beta$-oxidation, and cellular response to oxidative stress. In neurodegenerative diseases, these pathways are often impaired\cite{olufunmilayo2023oxidative}, leading to energy failure and excitotoxicity. In contrast, \texttt{Ocular Melanoma} exhibits enhanced $\beta$-oxidation\cite{lumaquin2023lipid}, supporting tumor proliferation and immune evasion. This inverse utilization of the same metabolic axis may reflect a mechanistic fork, shaped by the shared neural crest origin of retinal and neural tissues\cite{castro2023melanoma}. AADAT's dual role in neural excitotoxicity and tumor immune regulation further supports this. 


Although no clinical relationship has been established between neurodegenerative disorders and \texttt{Ocular Melanoma}, the observed transcriptomic similarities may reflect a shared developmental or metabolic context. These findings raise the possibility of underlying molecular features that span traditionally unrelated disease categories, which may merit further investigation through functional or mechanistic studies.

\subsection{Disease–Condition Interaction Analysis: Triggers of Comorbidity}

While disease–disease similarities often reflect shared genetic programs or pathological mechanisms, the presence of specific physiological or environmental conditions can further modulate the expression of such relationships. In our analysis, we explored disease–condition pairs to understand how background factors—such as obesity or hypertension—may shape transcriptomic overlaps and increase the risk of co-occurrence. 

One interesting case involves the co-occurrence of \texttt{Celiac Disease} and \texttt{Uterine Carcinosarcoma} in obese individuals. Though one is an autoimmune enteropathy and the other a rare uterine malignancy, they share dysregulation of genes such as \textit{A1CF}, \textit{AACS}, and \textit{ABCB1}, which point to altered mRNA editing, amino acid metabolism, and xenobiotic transport. These shared genes are enriched in pathways that are particularly sensitive to metabolic dysregulation in obesity, including glycosphingolipid biosynthesis, bile secretion, and branched-chain amino acid degradation. In obese individuals, chronic inflammation, disrupted metabolic homeostasis, and impaired detoxification mechanisms may jointly promote both autoimmune activation and tumorigenesis, thus creating a fertile biological landscape for comorbidity.

Another example is the comorbidity of \texttt{Acute Myeloid Leukemia (AML)} and \texttt{Osteoarthritis (OA)} in individuals with hypertension. These diseases converge on genes such as \textit{A2ML1} and \textit{A2M}, which regulate extracellular matrix homeostasis and inflammation, as well as \textit{A1CF}, which modulates immune signaling through RNA editing. The two diseases also share enrichment in immune and complement pathways, ECM degradation, and glucocorticoid response—features that are frequently exacerbated in hypertensive individuals. Hypertension, by promoting systemic inflammation, endothelial dysfunction, and hormonal imbalance, may amplify shared transcriptomic vulnerabilities in both hematologic and joint tissues.

This is not an isolated observation. In fact, 11 of the top 20 highest-scoring disease–condition pairs in our transcriptomic similarity analysis involve hypertension, including connections with \texttt{Acute Myeloid Leukemia, Adrenocortical Cancer, Gaucher Disease}, and \texttt{Osteoarthritis}. These findings underscore the wide-reaching systemic impact of hypertension—not just as a cardiovascular risk factor, but as a molecular amplifier of disease vulnerability across diverse biological systems. Given its high prevalence and silent progression, we emphasize the importance of early detection and integrative management of hypertension to mitigate its far-reaching comorbidity burden.

\subsection{Hypotheses Related to Rare Diseases}

\begin{figure}[ht]
\centering
\includegraphics[width=0.8\textwidth]{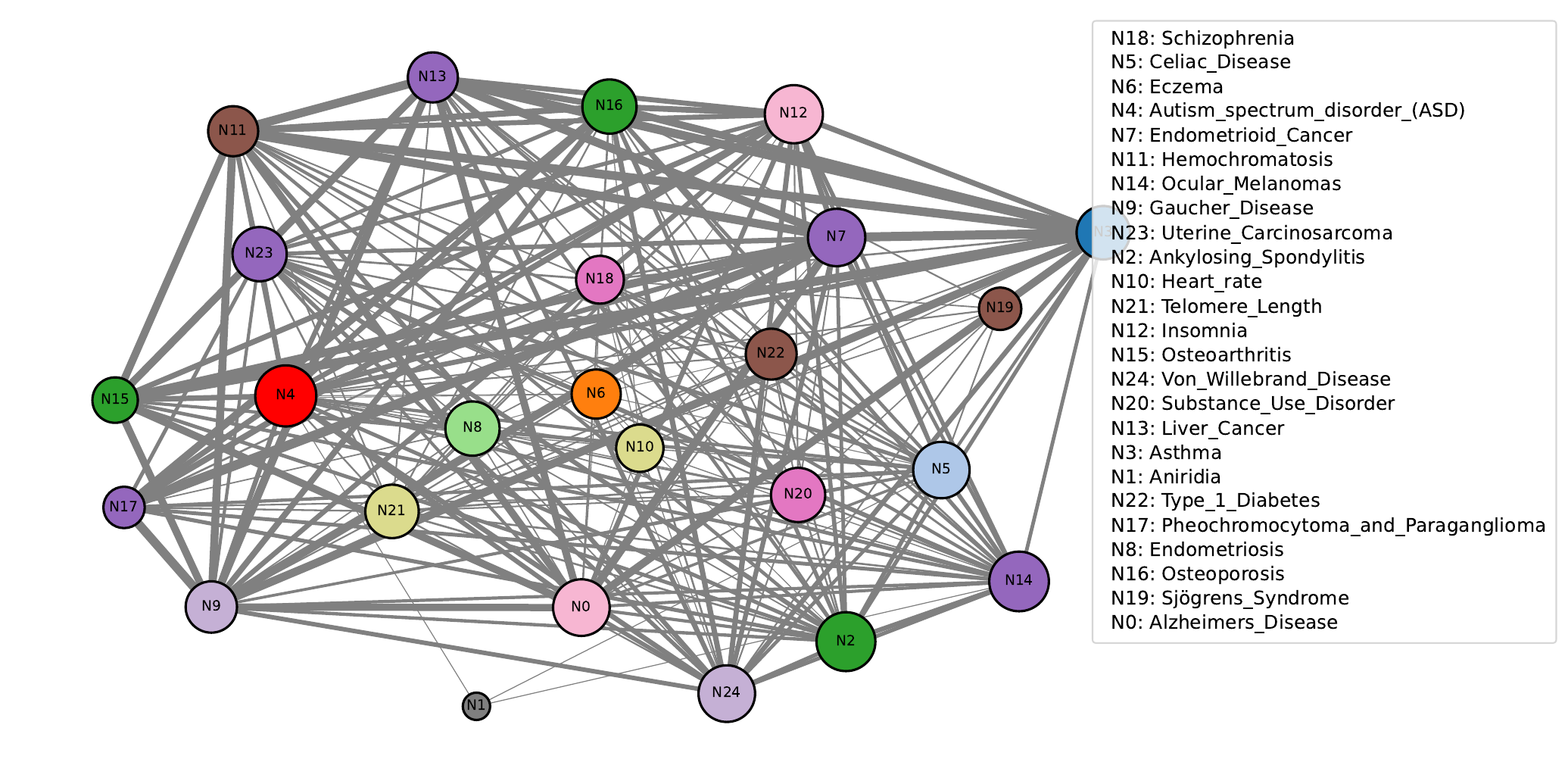}
\caption{
Subnetwork of diseases showing significant pathway-level similarity to ASD (the red node with index N4). Osteoporosis and Type 1 Diabetes (T1D) emerge as strongly connected conditions, both with established pharmacological pipelines.
}
\label{fig:asd_subgraph}
\end{figure}

In addition to mapping disease–disease similarity, we also examined whether transcriptomic overlaps with well-characterized conditions could highlight underexplored disorders that worth further investigation. For example, we extracted the subgraph centered on Autism Spectrum Disorder (ASD), a well-studied neurodevelopmental condition. As shown in Figure~\ref{fig:asd_subgraph}, this local network reveals close transcriptomic and pathway-level similarity between ASD and other conditions, including Osteoporosis and Type 1 Diabetes (T1D). While these two diseases are typically studied in distinct clinical domains, their established therapeutic pipelines and shared molecular features with ASD raise the possibility of identifying underexamined connections or therapeutic hypotheses, particularly in individuals with overlapping metabolic or immune phenotypes.

In the case of Osteoporosis, ASD shares genes such as \textit{AADAC}, \textit{ABCF3}, and \textit{ABCA7}, which participate in lipid metabolism and ABC transporter pathways. While these genes have not been directly targeted in ASD, several lipid-modulating agents—such as statins, bisphosphonates, and ANGPTL3/APOC3 inhibitors—have demonstrated activity along the same pathways. Their mechanistic action on lipid regulation and inflammatory balance raises the possibility that they could be repurposed for ASD, particularly in individuals with lipid signaling or neuroinflammation phenotypes.

A similar pattern emerges with T1D: shared genes like \textit{AADAT}, \textit{ABCD1}, and \textit{AATF} point to convergence in fatty acid oxidation, mitochondrial stress, and immune dysregulation. Corresponding therapeutic approaches—ranging from anti-inflammatory agents (e.g., $\alpha$1-antitrypsin, TYK2 inhibitors) to metabolic modulators (e.g., \textit{ABCD1} gene therapy, PPAR$\gamma$ agonists like Leriglitazone)—may offer a foundation for exploratory ASD interventions aimed at metabolic or immune correction.

While our findings are exploratory, they highlight potential directions for drug repurposing in ASD. We hope these cross-disease molecular links may serve as a starting point for future research into novel therapeutic strategies.

\section{Methodology}

\subsection{Large-Scale Gene Analysis via Transcriptomic Agentic AI System}
To explore disease relationships from a transcriptomic perspective, we leveraged the GenoTEX dataset \cite{liu2024genotex}, a large-scale, biologically curated benchmark for automated gene expression analysis. GenoTEX comprises 1,384 gene–disease association problems, spanning 132 distinct human diseases, each analyzed under varying biological or demographic conditions (e.g., age, sex, obesity, hypertension, or comorbidities). For clarity, we refer to the combination of a disease and a condition as a ``disease–condition pair", or simply a ``pair", throughout this study. The dataset encompasses 911 unique cohorts, totaling over 150,000 biological samples, with each cohort containing more than 18,000 normalized gene features on average.

To process this data at scale, we employed GenoMAS \cite{liu2025genomas}, a multi-agent agentic AI framework built for code-level automation in genomic analysis. 

Using this system, we performed end-to-end gene significance analysis for all 1,384 disease–condition pairs. 
The results include gene-level effect sizes (regression coefficients), lists of significant genes.

\subsection{Gene-Based Similarity Network}\label{method:gene_similarity}

To quantify transcriptomic similarity, we assessed gene overlap significance for each of the pairwaise combinations of 1384 pairs. For each pair, we retained genes with $|\beta| > 0.05$ from Lasso regression, filtering out weak associations. We then computed shared genes between each combination and performed a bidirectional hypergeometric test to evaluate whether the overlap exceeded random expectation, accounting for gene set sizes and the full gene universe (18,000+ genes). Benjamini–Hochberg correction was applied to adjust for multiple testing.

We retained only combinations with false discovery rate (FDR) $\leq 0.05$, yielding approximately 65,000 significant pairwise links out of nearly 1 million tested combinations. 

To avoid redundancy, we further filtered out biologically overlapping combinations involving generalized disease entries labeled as \texttt{None} (i.e., entries not conditioned on any specific biological factor). For instance, if a combination between \texttt{Disease1--None} and \texttt{Disease2--Obesity} was already significant, then additional combinations between \texttt{Disease1--Sex}, \texttt{Disease1--Age}, etc., and \texttt{Disease2--Obesity} were considered redundant and removed. This de-duplication step ensures that generalized associations do not inflate or obscure condition-specific links. There are 1293 links left after redundancy removal.

We constructed a similarity network using \texttt{NetworkX}, where each node represents a disease–condition pair and each edge denotes a statistically significant shared-gene relationship between two pairs. The edge weight was defined as $-\log_{10}(\text{FDR})$ to reflect the strength of transcriptomic similarity.

For visualization, we employed the \texttt{nx.spring\_layout()} algorithm, which arranges nodes in 2D space such that strongly connected nodes are pulled closer together. This layout reflects both the local and global structure of the similarity network.

To annotate the biological identity of each node, we assigned an ICD-10-CM category to every disease using GPT-4o. These categories were also used as node colors in the visualization for pathway-based similarity network (Figure~\ref{fig:pathway_net}).

\subsection{Pathway-Based Disease Similarity}\label{method:pathway_similarity}

To investigate functional overlap among disease–condition pairs, we performed pathway enrichment analysis and similarity scoring for all 1,293 significant combinations identified in the gene-level analysis.

\paragraph{Pathway Enrichment per Pair.}  
For each disease–condition pair, we mapped significantly expressed genes (with $|\beta| > 0.05$) to pathways using six complementary annotation databases: GO:Biological Process (GO:BP)\cite{gaudet2017go}, Reactome (REAC), KEGG\cite{kanehisa2002kegg}, transcription factor targets (TF), miRNA targets (MIRNA), and Human Phenotype Ontology (HP)\cite{kohler2019hpo}. This ensures both biological breadth and redundancy reduction.

We adopted the g:Profiler\cite{raudvere2019gprofiler} framework to retrieve enriched terms. g:Profiler selects higher-level, abstracted pathway terms to mitigate semantic variability across databases and maximize interpretability.

\paragraph{Identification of Shared Pathways.}  

For each disease–condition pairwise combination (hereafter, “combination”), we focused only on the genes that were shared between the two pairs. For each such shared gene, we retrieved its pathway annotations in both pairs. A pathway was considered “shared” if the same pathway was significantly enriched for the same gene in both pairs.

\paragraph{Similarity Scoring.}  

To quantify the overall strength of pathway-level similarity between two disease–condition pairs, we computed a cumulative score across all shared pathways using the following formula:
\[
\text{Similarity Score} = \sum_{k=1}^{N} \left[ \log(1 - p_{1k}) + \log(1 - p_{2k}) \right]
\]
\\
Here, \(N\) is the number of shared pathways between the two pairs, and \(p_{1k}\), \(p_{2k}\) are the enrichment p-values of the \(k\)-th pathway in the first and second pair, respectively. For each pathway, this score reduces to \(\log((1 - p_{1k}) \times (1 - p_{2k}))\), which reflects the joint probability that both enrichments are non-random. That is, the score becomes more positive when both \(p_1\) and \(p_2\) are small, indicating that the pathway is likely involved in both disease contexts. Summing across all such shared pathways allows us to capture not just the presence of overlap, but the joint confidence in their functional relevance.

\paragraph{Filtering.}  
We retained only combinations where at least one shared pathway had a positive similarity score, indicating non-random co-enrichment. This yielded 1,060 pathway-supported combinations out of the original 1,293 gene-sharing ones.

\paragraph{Pathway-level Graph Construction}
To visualize cross-disease functional similarity, we constructed an undirected weighted network in which each node represents a disease-condition pair. An edge is drawn between two nodes if the two diseases share at least one significantly enriched pathway. The edge weight corresponds to the pathway-level similarity score. Node size reflects its degree (i.e., the number of connected neighbors), and node color encodes ICD-10-CM categories. Similar to the gene-level network, we applied a spring-force layout in which edge lengths are inversely proportional to similarity weights. To better illustrate the strength of functional similarity, edge thickness is scaled proportionally to the similarity score—stronger similarities are rendered as thicker connections. 

In subsequent analyses, we also constructed disease-level subgraphs by collapsing the condition dimension (e.g., Figure~\ref{fig:asd_subgraph}). In these subgraphs, node size reflects the average degree of each disease across all associated conditions, while edge width and length correspond to the average pathway similarity score between the connected diseases.

\section{Discussion}



We envision several ways in which these resources may support future research. First, biomedical researchers may use the similarity network to formulate novel hypotheses about disease etiology, comorbidity, or molecular mimicry. For example, investigators studying a rare disease may identify common molecular patterns with a more prevalent condition, providing new leads for drug repurposing or biomarker development. Second, our gene and pathway-level results may help guide experimental validation studies, particularly in assessing the functional relevance of shared molecular programs. Finally, by providing structured, interpretable outputs from AI-generated analyses, we aim to lower the barrier for translational researchers to engage with complex transcriptomic datasets.

Importantly, many our findings are hypothesis-generating rather than confirmatory. While we identify plausible shared mechanisms, validation and causal inference remain critical next steps. We encourage researchers to treat these results as a foundation for targeted studies using additional data or experimental models.

More broadly, we advocate for stronger integration between AI systems and biomedical research—not just automation, but output that is biologically interpretable and clinically useful. We believe agentic AI, when grounded in biological context, can help bridge the gap between large-scale computation and biologically meaningful interpretation.

\section{Conclusion}

This study presents a transcriptomics-driven framework for rethinking disease relationships beyond traditional clinical boundaries. We propose a new similarity framework based on gene and pathway-level convergence, enabled by large-scale analysis of 1,300+ disease–condition pairs using the GenoMAS agentic AI system.

Our multi-layered analysis—from gene-level overlap to pathway-based convergence—captures both known comorbidities and novel cross-category links. We propose plausible molecular links across diseases, inferred from overlapping pathways related to metabolism, immune response, and cellular stress. Systemic conditions like obesity and hypertension emerge as modulators of transcriptomic similarity, while rare diseases such as autism spectrum disorder may benefit from therapeutic insights derived from better-characterized conditions.

By sharing our full results and network resources, we hope to support hypothesis generation, validation, and translational research. More broadly, this work demonstrates how agentic AI can act as a biologically informed collaborator—scaling analysis while enabling mechanistic interpretation across complex disease landscapes.

\bibliographystyle{ws-procs11x85}
\bibliography{ws-pro-sample}

\end{document}